\documentclass[apjl]{emulateapj}

\usepackage{color}

\newcommand{\FLASH}{{\sc flash}}

\slugcomment{Submitted to the \apjl}

\shorttitle{Gravitationally Confined Detonation}
\shortauthors{Plewa et~al.}

\begin{document}

\title{Type Ia Supernova Explosion: Gravitationally Confined Detonation}

\author{
T. Plewa\altaffilmark{1,2,3},
A. C. Calder\altaffilmark{1,2}, and
D. Q. Lamb\altaffilmark{1,2,4}
}

\altaffiltext{1}{Center for Astrophysical Thermonuclear Flashes,
   The University of Chicago,
   Chicago, IL 60637}
\altaffiltext{2}{Department of Astronomy \& Astrophysics,
   The University of Chicago,
   Chicago, IL 60637}
\altaffiltext{3}{Nicolaus Copernicus Astronomical Center,
   Bartycka 18,
   00716 Warsaw, Poland}
\altaffiltext{4}{Enrico Fermi Institute,
   The University of Chicago,
   Chicago, IL 60637}

\begin{abstract}

We present a new mechanism for Type Ia supernova explosions in massive
white dwarfs.  The proposed scenario follows from relaxing the
assumption of symmetry in the model and involves a detonation created
in an unconfined environment. The explosion begins with an essentially
central ignition of stellar material initiating a deflagration. This
deflagration results in the formation of a buoyantly-driven bubble of
hot material that reaches the stellar surface at supersonic
speeds. The bubble breakout forms a strong pressure wave that
laterally accelerates fuel-rich outer stellar layers.  This material,
confined by gravity to the white dwarf, races along the stellar
surface and is focused at the location opposite to the point of the
bubble breakout. These streams of nuclear fuel carry enough mass and
energy to trigger a detonation just above the stellar surface. The
flow conditions at that moment support a detonation that will
incinerate the white dwarf and result in an energetic explosion. The
stellar expansion following the deflagration redistributes stellar
mass in a way that ensures production of intermediate mass and
iron group elements consistent with observations. The ejecta will have
a strongly layered structure with a mild amount of asymmetry following
from the early deflagration phase. This asymmetry, combined with the
amount of stellar expansion determined by details of the evolution
(principally the energetics of deflagration, timing of detonation, and
structure of the progenitor), can be expected to create a family of
mildly diverse Type Ia supernova explosions.

\end{abstract}


\section{Introduction}
Type Ia supernovae are one class of luminous stellar explosions. These
are the predominant explosive events in old stellar environments such
as elliptical galaxies. The ejecta of these objects are rich in
intermediate mass and iron group elements. Explaining the nature of
these objects is therefore critical for understanding galactic
chemical evolution \citep{truran+71}.  These supernovae also are the
key component of one method used to determine the history of the
Universe and probe the origin of dark energy
\citep{sandage+93,perlmutter+99,tonry+03,knop+03}.

The work presented in this {\em Letter} builds on many previous
observational and theoretical contributions to the field of Type Ia
SNe, and on advances in fluid dynamics, nuclear physics, and
computational science.  Current ideas about the Type Ia supernova
explosion mechanism follow from the original work of Arnett, Nomoto,
and Khokhlov \citep{arnett69,nomoto+76,khokhlov91-dd}, who pioneered
deflagrating and detonating massive white dwarfs originally proposed
by \cite{hoyle+60} as the core component of Type Ia supernovae.

Despite decades of effort, these events remain an unsolved mystery.
Proposed explosion scenarios include white dwarf detonations
\citep{arnett69,nomoto82}, coalescing pairs of white dwarfs
\citep{webbink84,iben+84}, deflagrations or delayed detonations of
massive white dwarfs \citep{nomoto+84,khokhlov91-dd}, and collapse in
a strong gravitational field \citep{wilson+04}. None of these
scenarios accounts for all the observed features of Type Ia
supernovae.  Some models produce energetic explosions but fail to
explain the observed ejecta compositions, while others successfully
produce the observed chemically stratified ejecta but require
including ad hoc physics.

Here we report the results of multi-dimensional hydrodynamical
simulations of the long-term evolution of a massive white dwarf
following an essentially central ignition. The initial evolution
results in a deflagration front and formation of a Rayleigh-Taylor
unstable buoyancy-driven bubble that is accelerated to supersonic
speeds on its way to the stellar surface.  We follow the evolution
beyond bubble breakout and observe formation of gravitationally
confined flow across the surface of the star. The flow is focused at
the point opposite the breakout location, where the matter is
compressed and heated, igniting a detonation wave just above the
stellar surface. We find that the star expands substantially during
the evolution.  This expansion produces a density distribution that,
when overrun by the detonation wave, can be expected to result in
ejecta that are strongly layered and rich in intermediate mass
elements, as is typical of Type Ia supernovae. The modest asymmetry in
the expansion of the star can also be expected to produce a mild
amount of global asymmetry in the explosion.
\section{Numerical model} 
The simulations presented here were performed with the adaptive mesh
refinement hydrodynamics code \FLASH\ \citep{fryxell+00}.  The
numerical scheme includes self-gravity solved using a multipole
expansion. We have established that to properly account for
asymmetries in the mass distribution and to ensure momentum
conservation in the model, at least $3$ multipole moments are needed
in the expansion.  In the simulations reported here, we used $10$
multipole moments.  The rest of the physics modules are identical to
those used in our previous study \citep{calder+04}, in particular the
evolution of the deflagration front is computed with a flame capturing
scheme with energy release accounting for carbon, magnesium, and
silicon burning \citep{khokhlov01}.

The computational domain is a two-dimensional region in cylindrical
geometry covering a region from $-16,384$~km to $16,384$~km in the
z-direction and extending up to $16,384$~km in radius.  We employed an
outflow-only boundary condition at all domain boundaries, except along
the symmetry axis at $r=0$ where we used a reflecting boundary
condition.  The adaptive mesh was used to track flow features at a
maximum effective resolution of 8~km (corresponding to an effective
grid resolution of $2048\times 4096$). The simulation was executed at
Courant number of $0.6$, and the time step was limited to a maximum of
$4\times10^{-4}$~s.

The initial conditions consist of a $1.36$ solar mass isothermal white
dwarf composed of equal amounts of carbon and oxygen with a
temperature of $3\times10^7$\,K. The progenitor model was mapped to
the computational grid following the procedure employed in
\cite{calder+04}. The nuclear flame was initiated as a small spherical
region of completely burned material placed in hydrostatic equilibrium
with its surroundings. The ignition region centered at
$(r,z)=(0,12.5)$~km had a radius of $50$~km.
\section{Results}
\begin{figure*}[ht]
\begin{center}
\includegraphics[height=7cm,clip=true]{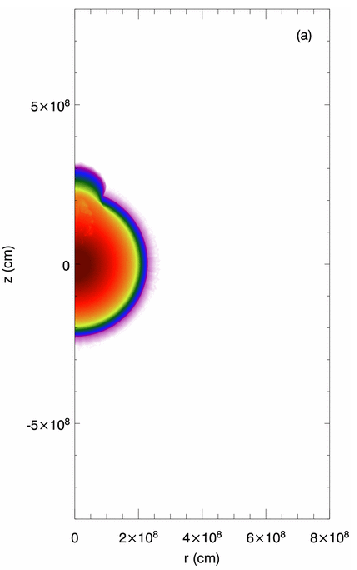}%
\includegraphics[height=7cm,clip=true]{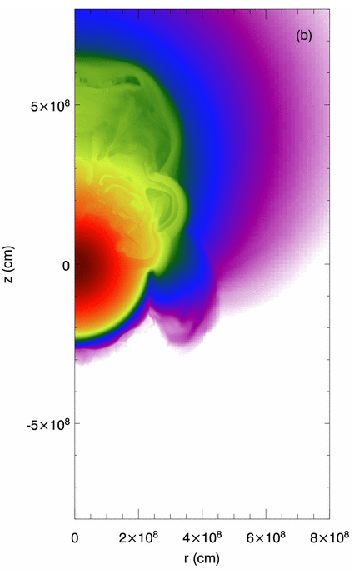}%
\includegraphics[height=7cm,clip=true]{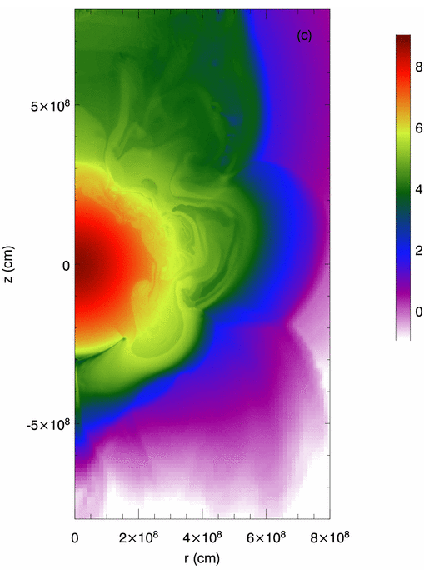}
\end{center}
\caption{Evolution of the white dwarf following an essentially central
ignition.  Density in log-scale is shown for the innermost
approximately one-sixth of the simulation domain. (a) The bubble
breakout ($t=0.9$~s). The material is expelled radially, and the high
pressure of the burned bubble material produces a lateral acceleration
of the outer layers of the star.
(b) The bubble material expands above the star as the fuel-rich surface
layers reach the equator in their race across the stellar surface
($t=1.4$~s). Note that most of the surface layers are closely confined
to the star.
(c) Focusing of the fuel-rich streams just prior to the detonation
($t=1.9$~s).  Note the presence of the dense conical region stretching
down along the symmetry axis beginning at $(r,z)=(0,-3\times 10^8)$~cm.
}
\label{f:density_sequence}
\end{figure*}

Figure \ref{f:density_sequence} depicts several moments in the evolution 
of the density of the star from the point of bubble breakout to just
prior to the ignition of the detonation.
The rapid ascent of the bubble toward the stellar surface accelerates
the stellar material located just above the bubble.  This piston-like
behavior results in the formation of a bulge filled with high-pressure
high-momentum nuclear ash.  During breakout, bubble material is
expelled mostly radially, while the high pressure of the burned
material accelerates the surface layers laterally
(Fig.~\ref{f:density_sequence}(a)). This material races along the
stellar surface, followed by the magnesium-rich bubble material.  Both
remain gravitationally confined to a relatively thick $\approx
1000$~km layer (Fig.~\ref{f:density_sequence}(b)). At $t\approx
1.8$~s, this flood of surface material converges at the point opposite
to the bubble breakout location, forming a conical compressed region
bounded by the shock. This structure stretches down the symmetry axis
beginning at $z\approx -3\times 10^8$~cm
(Fig.~\ref{f:density_sequence}(c)).

\begin{figure*}[ht]
\begin{center}
\includegraphics[width=7cm,clip=true]{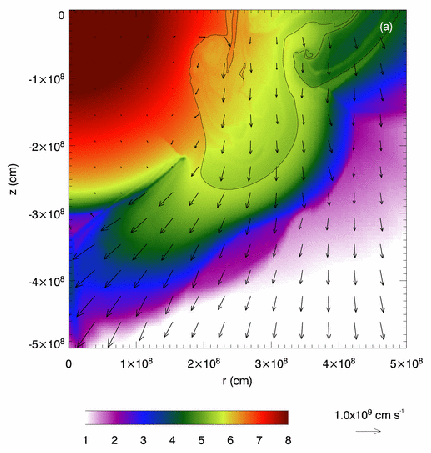}%
\includegraphics[width=7cm,clip=true]{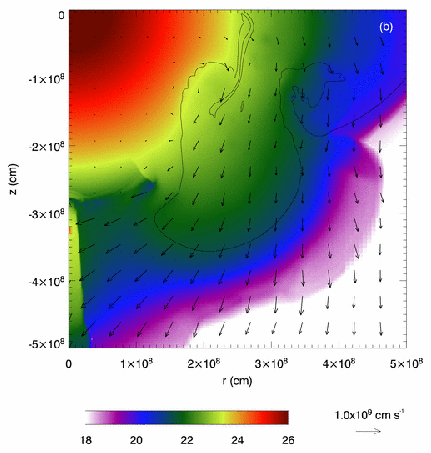}
\end{center}
\caption{Evolution of the surface flood following an essentially
central ignition of the white dwarf. The lower half of the star is
shown in the figures.
(a) Density in log-scale at $t= 1.85$~s. Note the density gradient
present inside the stream of fast-moving material flowing across the
stellar surface.  Also note that the interior of the star remains
highly radially symmetric, and that only relatively low-density regions
are perturbed by the surface flow.
(b) Pressure in log-scale at the time of the ignition of the detonation
($t= 2.005$~s). The ignition point can be seen as a small over-pressured
region located near the symmetry axis at $(r,z)\approx (0,-3.25\times
10^8$)~cm. The shock wave preceding the bubble material can also be
seen in the low-density stellar layers near $(r,z)\approx (1.1\times
10^8,-2.4\times 10^8)$~cm. The contour marks the position of the
advancing front of burned bubble material.  Vectors show the velocity
field; the magnitudes of these velocities can be determined by
comparing the length of these vectors  with the length of the fiducial
vector, which corresponds to $1\times 10^9$~cm~s$^{-1}$.
}
\label{f:detail}
\end{figure*}

At this point in the evolution, conditions in the shocked region
approach the detonation regime.  By $t=1.85$~s, material upstream of
the confluence region has density of $\approx 10^4$~g~cm$^{-3}$ and
moves at velocity $9,500$~km~s$^{-1}$. Downstream of the shock, the
density of the nuclear fuel is $\approx 5\times 10^4$~g~cm$^{-3}$ and
the temperature reaches $\approx 1.4\times 10^9$~K.  Due to a mild
density gradient present in the flooding material
(Fig.~\ref{f:detail}(a)),
the post-shock density slowly increases with time while the
temperature remains relatively constant. At $t\approx 1.93$~s, the
post-shock conditions are suitable for igniting the nuclear fuel: the
density exceeds $1.7\times 10^6$~g~cm$^{-3}$ and the temperature is
$\approx 2.2\times 10^9$~K. The detonation point can be seen as a
slightly over-pressured region located near the symmetry axis at
$(r,z)\approx (0,-3.35\times 10^8$)~cm (Fig.~\ref{f:detail}(b)).

The detonation wave born in the region above the stellar radius will
sweep through the white dwarf, which underwent a substantial evolution
from the moment of the ignition of the deflagration.  During its
ascent, the rising bubble displaced about 5\% of the stellar mass.
This mass displacement, combined with the pressure wave caused by
nuclear energy release, leads to expansion of the star.  Thermal
expansion is present from the onset of the deflagration with that part
of the star nearest the deflagration experiencing relatively stronger
thermal expansion. The global evolution of the stellar matter is,
however, primarily in response to the softening of the gravitational
potential caused by mass displacement.  This expansion has a mostly
radial character and becomes effective only after bubble breakout.

\begin{figure}[ht]
\begin{center}
\includegraphics[width=8cm,clip=true]{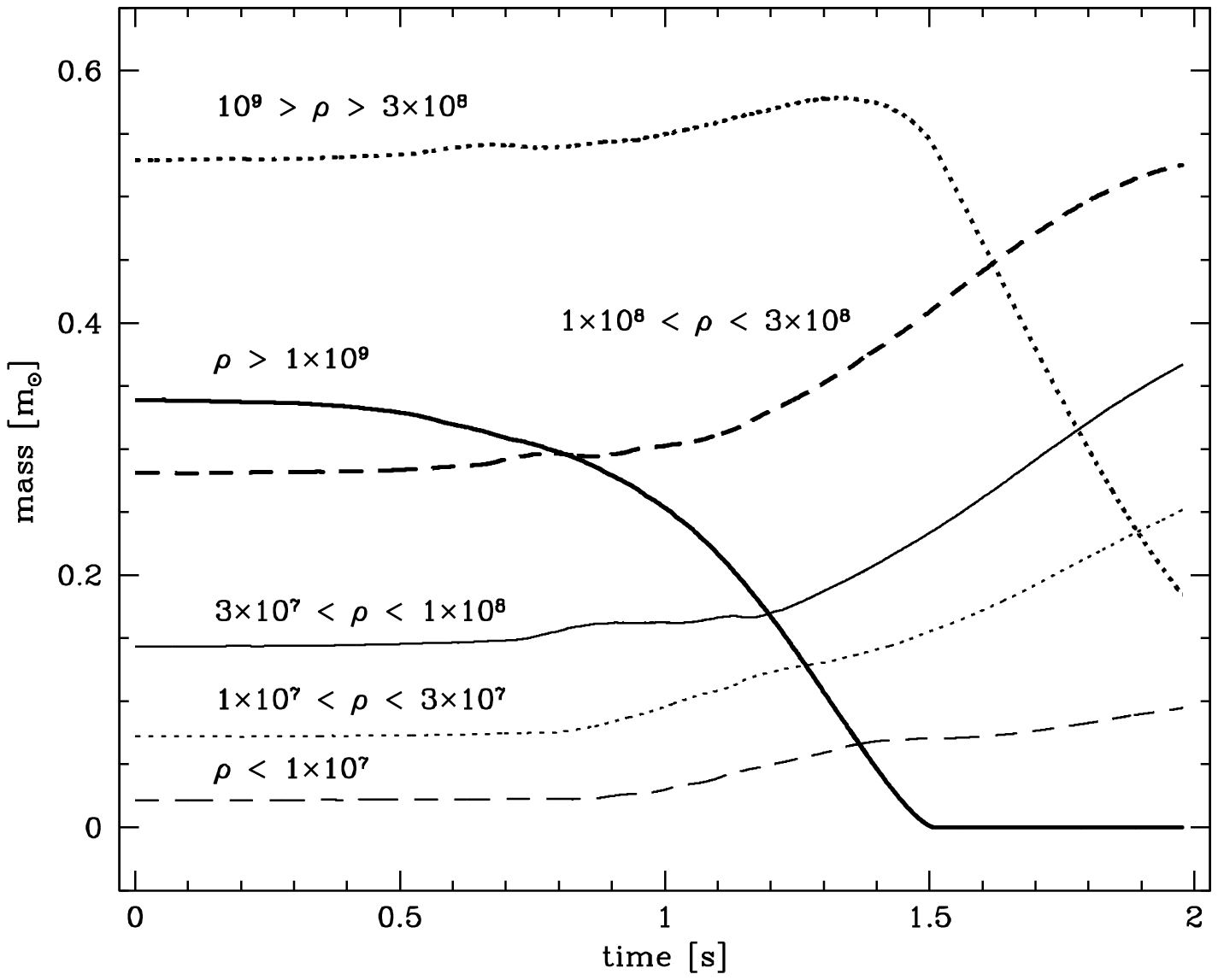}
\end{center}
\caption{Evolution of the mass distribution in the model white dwarf.
Amount of material in solar masses inside select density intervals is
shown as a function of time.  Significant stellar expansion takes
place only after bubble breakout (about $0.9$~s after the deflagration
began near the center of the star). The expansion velocity increases
linearly from the stellar center and leads to a steady decrease in the
density of the star at all radii.  At the moment of the ignition of
the detonation, almost half of the stellar mass has densities below
$1\times 10^8$~g~cm$^{-3}$.  }
\label{f:mass_distribution}
\end{figure}

Figure \ref{f:mass_distribution}
shows the evolution of density during the course of the simulation.
Shown is the amount of mass in several density intervals.  The
material with density $>1\times 10^8$~g~cm$^{-3}$ will undergo burning
to iron peak elements, while material with density below $3\times
10^7$~g~cm$^{-3}$ will burn into intermediate mass elements. Several
points can be noted here. There is almost no stellar expansion prior
to the bubble breakout ($t\leq 0.9$~s). The breakout is followed by
relatively fast expansion of the densest material ($\rho > 1\times
10^9$~g~cm$^{-3}$), which is succeeded by uniform expansion
(identified by a simultaneous increase in the expansion rates at
$t\approx 1.2$~s). This process continues until the ignition of the
detonation, with the most rapid expansion at densities between
$1\times 10^7$~g~cm$^{-3}$ and $1\times 10^8$~g~cm$^{-3}$.  At the
time of detonation the amount of mass at the densities characteristic
for production of intermediate mass elements, $\rho < 3\times
10^7$~g~cm$^{-3}$, is 0.35~$M_\sun$.  The amount of mass at densities
above $1\times 10^8$~g~cm$^{-3}$, required for production of the iron
peak elements, is 0.71~$M_\sun$.
\section{Discussion and Conclusions}
We presented a gravitationally confined detonation (GCD) mechanism for
Type Ia supernova explosions. The basic components of the proposed
scenario are a rising deflagrating bubble expelling a small amount of
stellar matter, the associated stellar expansion caused by the
shallower potential well, the flood of stellar material across the
surface following the bubble breakout, and a detonation in an
gravitationally confined environment.

In the GCD scenario, not imposing any asymmetries in the initial
conditions is of paramount importance. A deflagration front is born
very close to the white dwarf center.  Such an essentially central
ignition is more probable than the idealized conditions adopted in
standard deflagration or delayed detonation models, primarily because
the central region of the star is convective.  This type of ignition
results in a rising deflagrating bubble accelerated by buoyancy to
supersonic speeds on its way to the stellar surface.  The transonic
phase of the bubble's rise is accompanied by the formation of a bow
shock ahead of the bubble that compresses and heats the nuclear fuel.
Our attempts to associate this region with a possible transition to
detonation failed. We discovered, however, that the flood of the
expelled surface layers that follows the bubble breakout remains
confined to low radii above the star. This flood races around the star
and is ultimately focused into a hot, compressed, high density region
located just above the stellar surface.  Conditions in this region
satisfy the criteria necessary for a detonation.

One important aspect of the GCD mechanism is that stellar expansion is
a natural consequence of the essentially central ignition of a
deflagration.  The flame releases energy and
displaces mass, softening the gravitational potential well leading to
expansion of the star.  The expansion will slow down on a time scale
comparable to the sound crossing time of the white dwarf as the star
tries to reach a new equilibrium.  Therefore, we expect that at still
later times, if not for the fact that the detonation will completely
disrupt the white dwarf, the initial expansion would be followed by
contraction of stellar material and the star would oscillate.

Because of this pre-expansion, the detonation front born above the
stellar surface will encounter densities similar to those found in
models where pre-expansion results from a centrally ignited
large-scale deflagration~\citep{reinecke+02,gamezo+03}. The estimate
of nucleosynthetic yield for intermediate mass (iron peak) elements is
a lower (upper) limit in view of the fact that the stellar expansion
continues after the moment of the detonation. Determining the final
yields, however, requires simulating the detonation. The actual
conditions across the detonation wave, particularly the amount of
compression, will influence the results.  Also, the yields can be
modified by delaying or by accelerating the ignition of the detonation
to create a family of Type Ia supernovae with diverse spectral
characteristics. The exact timing of the detonation depends on several
factors. The structure of the progenitor influences energy release by
the deflagration. It also affects the strength and mass of the stellar
layers being pushed around the star. The radius of the progenitor
regulates the time required by the streaming matter to reach the
confluence point. All these factors determine the time available for
stellar pre-expansion and will be a source of diversity in GCD models.

Some properties of the proposed model are, however, largely
independent of the precise details of the ignition of the detonation
or stellar progenitor. The explosion will be powerful. All the stellar
fuel will be consumed by the detonation.  Despite the perturbation
introduced by the deflagration, the star will retain most of its
radially symmetric stratification. Therefore, the subsequent explosion
will display characteristics well-known from one-dimensional
investigations \citep{nomoto+84,hoeflich+98}. In particular, we expect
the distribution of nucleosynthetic products in velocity space to
agree with the observed layered structure of Type Ia supernova ejecta.

The model also naturally admits certain asymmetries. The deflagration
consumes only about 5\% of the stellar mass. We expect a similar level
of variation in the resulting spectra and luminosities, in agreement
with degree of diversity present in the observations \citep{li+01}. In
addition, because the stellar shape is distorted by the rising bubble
and the formation of the detonation on one side of the star, we expect
a noticeable asymmetry in the stellar ejecta. These orientation
effects might be responsible for peculiar events such as SN 1991T
\citep{filippenko97}.  Because the gross properties of observed Type
Ia supernovae can be accounted for in the GCD model, detailed
observations will be required to verify the proposed mechanism.  One
possibility is to obtain information about the degree of asymmetry in
these explosions, using precision polarimetric measurements.

In this short communication, we have not presented simulations that
support all of our predictions.  Rather, we have presented a logical
sequence of events that naturally leads to a Type Ia supernova
explosion. Several of these steps need to be carefully studied. In
particular, the proposed detonation mechanism bears many similarities
to the process of confined fusion studied in terrestrial laboratory
experiments, which is notoriously difficult and prone to
instabilities.  For this reason, the early stages of the detonation
should be studied very carefully.  The expected computational demands
and challenges are severe and clearly approach limits of feasibility.
Despite the fact that such a study lies in the future, we are
confident in the basic components of the proposed GCD mechanism.
\acknowledgements
The authors thank  Alexei Khokhlov and Peter H\"oflich for their
comments. Contributions from Natasha Vladimirova, Ed Brown, Jim
Truran, and the FLASH Code Group made this work possible. This work is
supported in part by the U.S. Department of Energy under Grant
No.\ B341495 to the Center for Astrophysical Thermonuclear Flashes at
the University of Chicago.

\end{document}